\documentclass{article}
\usepackage{ijcai18}

\usepackage{times}
\usepackage{xcolor}
\usepackage{soul}
\usepackage[utf8]{inputenc}
\usepackage[small]{caption}
\usepackage{epsfig}
\usepackage{graphicx}
\usepackage{subfigure}
\usepackage{amsmath}
\usepackage{amssymb}
\usepackage{multirow}
\usepackage{algorithmic}
\usepackage{algorithm}
\usepackage{array}

\title{Cross-media Multi-level Alignment with Relation Attention Network}

\author{Jinwei Qi, Yuxin Peng\thanks{Corresponding author.} and Yuxin Yuan\\ 
Institute of Computer Science and Technology, Peking University, Beijing 100871, China  \\
pengyuxin@pku.edu.cn}

\begin{document}

\maketitle

\begin{abstract}
	With the rapid growth of multimedia data, such as image and text, it is a highly challenging problem to effectively correlate and retrieve the data of different media types.
	Naturally, when correlating an image with textual description, people focus on not only the alignment between discriminative image regions and key words, but also the relations lying in the visual and textual context. Relation understanding is essential for cross-media correlation learning, which is ignored by prior cross-media retrieval works. To address the above issue, we propose Cross-media Relation Attention Network (CRAN) with multi-level alignment.  
	First, we propose \textbf{visual-language relation attention model} to explore both fine-grained patches and their relations of different media types. We aim to not only exploit cross-media fine-grained local information, but also capture the intrinsic relation information, which can provide complementary hints for correlation learning.
	Second, we propose \textbf{cross-media multi-level alignment} to explore global, local and relation alignments across different media types, which can mutually boost to learn more precise cross-media correlation. 
	We conduct experiments on 2 cross-media datasets, and compare with 10 state-of-the-art methods to verify the effectiveness of proposed approach.
\end{abstract}

\section{Introduction}

\begin{figure}[!t]
	\centering
	\includegraphics[width=0.49\textwidth]{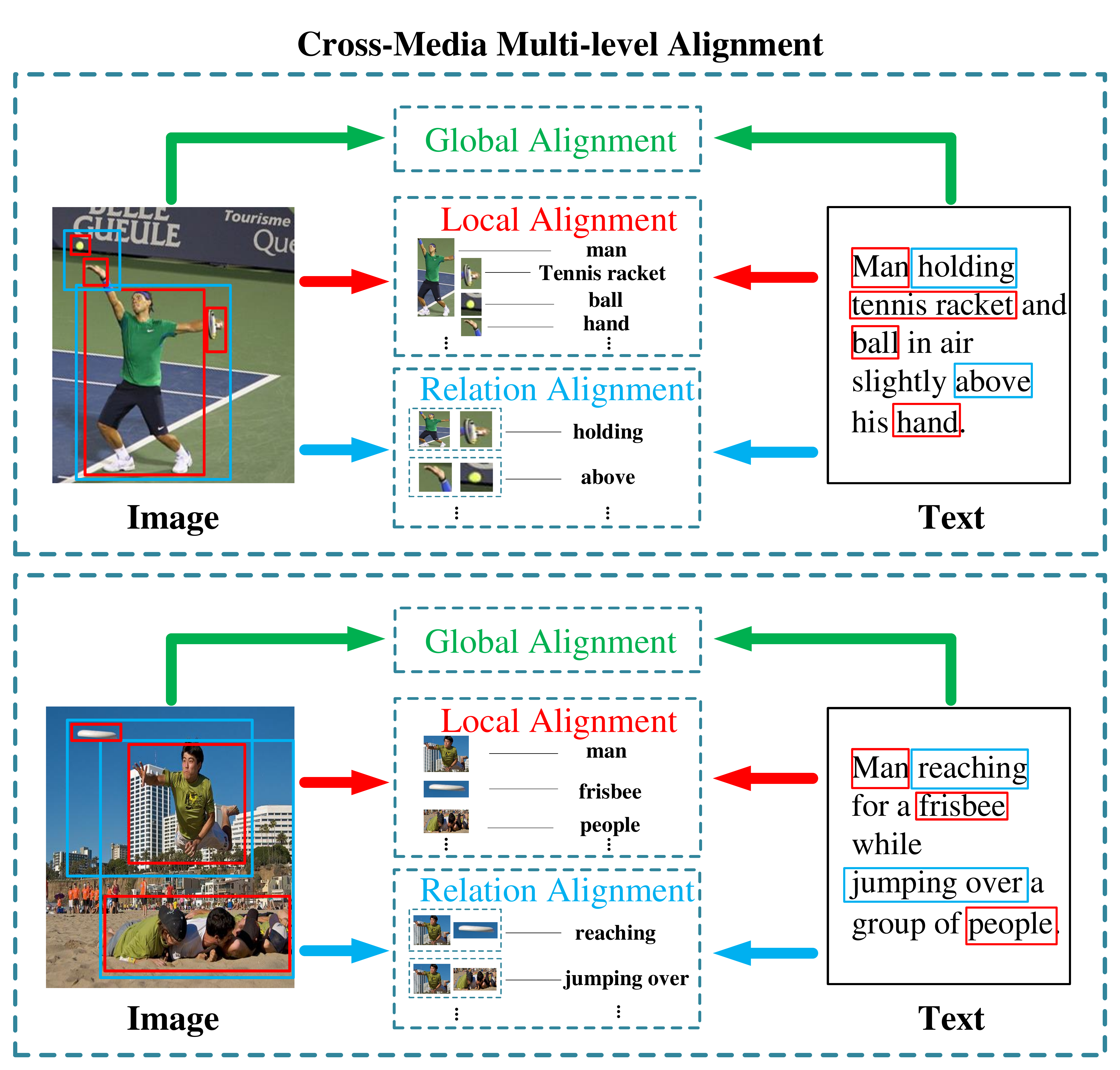}
	\caption{An example of cross-media multi-level alignment for correlation learning, which not only explores global alignment between original instances and local alignment between fine-grained patches, but also captures relation alignment lying in the context.}
	\label{fig_cross_media}
\end{figure}

Nowadays, multimedia data has been emerging rapidly on the Internet, including image and text. 
Under these circumstances, cross-media retrieval has become an essential technique for search engine as well as multimedia data management. It can provide flexible retrieval experience to search the data of different media types simultaneously by a query of any media type. 
However, ``heterogeneity gap'' among multimedia data causes inconsistent distributions and representations of different media types, which makes it quite challenging to effectively measuring cross-media similarity. 

Cross-media correlation naturally exists among the heterogeneous data with latent semantic alignment, and the research of cognitive science further indicates that in human brain, cross-media correlation can be fully understood through the fusion of multiple sensory organs, such as visual and language mechanism. Furthermore, when matching an image with textual description, people not only align the key words with discriminative image regions, but also consider the alignment of context information, which is represented by relations between image regions as well as their language expression. Therefore, there exists multi-level alignment between image and text, as shown in Figure \ref{fig_cross_media}. It is necessary to fully exploit and understand latent cross-media correlation from various aspects, and construct metrics on the data of different media types, which has been the key problem of cross-media retrieval.

For bridging the ``heterogeneity gap'', most of existing methods follow the intuitive idea to model cross-media correlation by constructing common space. The features of different media types are projected into common space, so that cross-media similarity can be directly measured between common representations. According to their different models, these methods can mainly be divided into two categories: The first is traditional methods, which maximize the correlation of variables from different media types. They mainly learn mapping matrices to construct common space. Canonical correlation analysis based methods \cite{RasiwasiaMM10SemanticCCA,DBLP:journals/neco/HardoonSS04,DBLP:journals/ijcv/GongKIL14} are the main stream, while others \cite{ZhaiTCSVT2014JRL} attempt to integrate graph regularization in correlation learning. But their performance is limited in traditional framework. Thus it leads to the second kind of methods. They utilize strong learning ability of deep neural network to model cross-media correlation. Such methods like \cite{feng12014cross,DBLP:conf/icml/AndrewABL13,peng2017ccl} construct multi-pathway network to learning common representation.

However, the aforementioned methods mainly explore cross-media correlation from either global original instances or local fine-grained patches. While they ignore the relations between different fine-grained patches. There can form multiple alignments, which indicate the correspondence on the context of different media types, and provide rich complementary hints for correlation learning. Therefore, it should be effectively considered. 
For addressing the above problem, we propose Cross-media Relation Attention Network (CRAN) with multi-level alignment, which has the following contributions.
\begin{itemize}
	\item \textbf{Visual-language relation attention model}. We utilize attention mechanism to explore not only the local fine-grained patches as discriminative image regions and key words, but also the relations in both visual and textual context. We aim to fully exploit fine-grained local information and intrinsic relation information, which can provide complementary hints for cross-media correlation learning.
	
	\item \textbf{Cross-media multi-level alignment}. We not only exploit global alignment between original instances and local alignment between fine-grained patches, but also mine relation alignment between the context of different media types, which can mutually boost to learn more precise cross-media correlation, and further promote cross-media retrieval.
	
\end{itemize}
To verify the effectiveness of our proposed CRAN approach, we conduct cross-media retrieval experiments on 2 widely-used datasets compared with 10 state-of-the-art methods.

\section{Related Works}

We briefly review representative methods of cross-media retrieval, which mostly perform cross-media correlation learning to construct common space. Thus, the similarity of heterogeneous data can be calculated on common representations. These methods can be divided into two categories, namely \textit{traditional methods} and \textit{deep learning based methods}.

\subsection{Traditional Methods}

Traditional methods mainly attempt to maximize the correlation between pairwise data of different media types. They learn mapping matrices to project multimedia data into common space, and generate common representations. Some representative methods utilize canonical correlation analysis (CCA) \cite{HotelingBiometrika36RelationBetweenTwoVariates} to optimize the statistical values. They construct a lower-dimensional common space, and there are many extensions based on them. 
For example, Rasiwasia et al. integrate semantic category labels with CCA to perform semantic matching \cite{RasiwasiaMM10SemanticCCA}. 
Gong et al. propose Multi-view CCA to construct a third view of high-level semantics \cite{DBLP:journals/ijcv/GongKIL14}.
While Multi-label CCA \cite{DBLP:conf/iccv/RanjanRJ15} considers semantic information with multiple label annotations. 
Similar to CCA, Li et al. propose cross-modal factor analysis (CFA) \cite{LiMM03CFA}, which learns projections by minimizing the Frobenius norm between pairwise data. 
Besides, some other traditional methods attempt to utilize graph regularization. 
They construct graphs for data of different media types, and preform cross-media correlation learning. 
Zhai et al. propose joint representation learning (JRL) to construct several separate graphs \cite{ZhaiTCSVT2014JRL}, and they also integrate semi-supervised information to learn common space. Peng et al. further construct a unified hypergraph to exploit fine-grained information simultaneously \cite{DBLP:journals/tcsv/PengZZH16}. 

\begin{figure*}
	\begin{center}
		\includegraphics[width=1.0\textwidth]{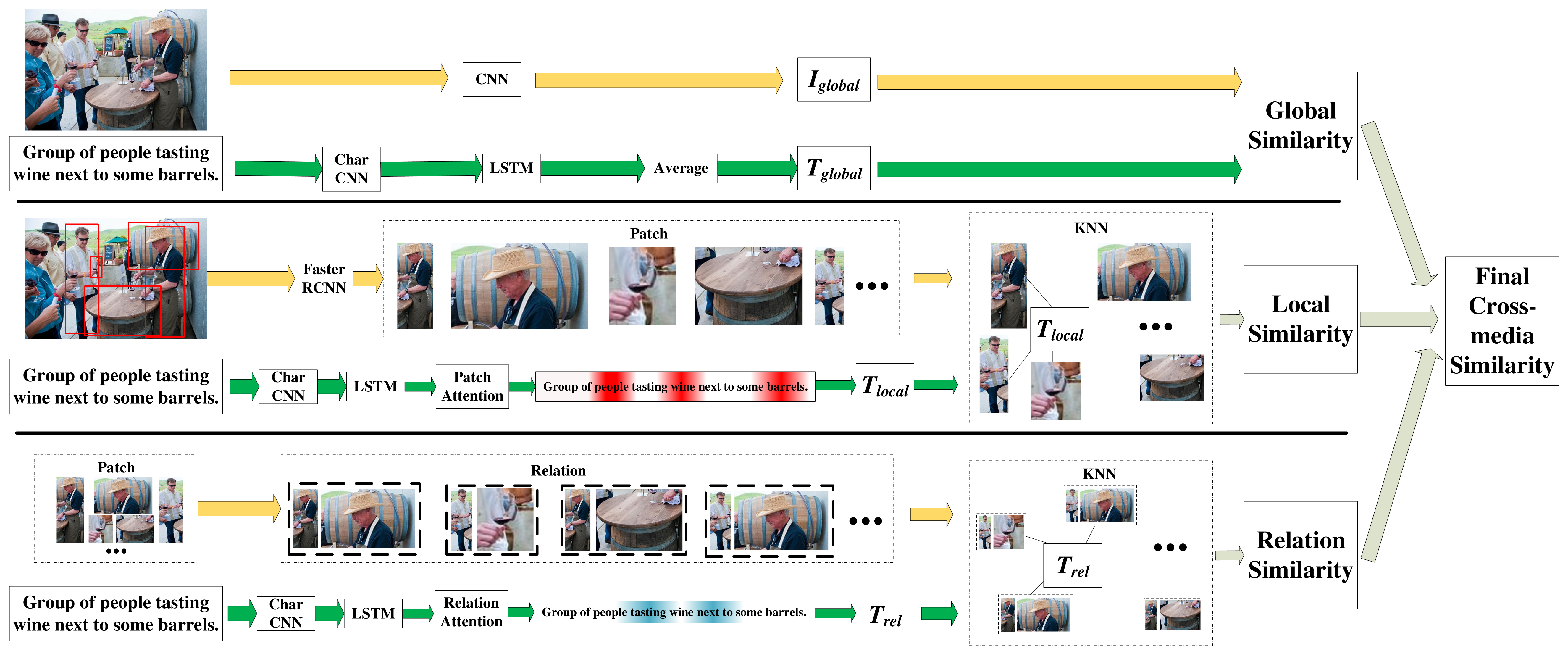}
	\end{center}
	\caption{An overview of our proposed CRAN approach. Multi-level alignment is fully exploited with relation attention network between the global original instances, local fine-grained patches as well as their relations, which aims to learn more precise cross-media correlation.}
	\label{fig:network}
\end{figure*}

\subsection{Deep Learning based Methods}

With the recent advances of deep learning in multimedia applications, such as image classification \cite{ImageNet2012} and object detection \cite{DBLP:conf/nips/RenHGS15}, researchers adopt deep neural network to learn common space for cross-media retrieval, which aims to fully utilize its considerable ability of modeling highly nonlinear correlation. 
Most of deep learning based methods construct multi-pathway network, where each pathway is for the data of one media type. 
	Multiple pathways are linked at the joint layer to model cross-media correlation. 
	Ngiam et al. propose bimodal autoencoders (Bimodal AE) to extend restricted Boltzmann machine (RBM) \cite{ngiam32011multimodal}. They model the correlation by mutual reconstruction between different media types. 
Multimodal deep belief network \cite{srivastava2012learning} adopts two kinds of DBNs to model the distribution over data of different media types, and it constructs a joint RBM to learn cross-media correlation.
Andrew et al. propose deep canonical correlation analysis (DCCA) to combine traditional CCA with deep network \cite{DBLP:conf/icml/AndrewABL13}, which maximizes correlation on the top of two subnetworks. Feng et al. jointly model cross-media correlation and reconstruction information to perform correspondence autoencoder (Corr-AE) \cite{feng12014cross}. 
Furthermore, Peng et al. propose cross-media multiple deep networks (CMDN) \cite{DBLP:conf/ijcai/PengHQ16}. They construct hierarchical network structure with stacked learning strategy, which aims to fully exploit both inter-media and intra-media correlation. 
	Cross-modal correlation learning (CCL) \cite{peng2017ccl} utilizes fine-grained information, and adopts multi-task learning strategy for better performance.

However, above methods mainly focus on pairwise correlation, which exists in global alignment between original instances of different media types. Although some of them attempt to explore local alignment between fine-grained patches, they all ignore important relation information lying in the context of these fine-grained patches, which can provide rich complementary hints for cross-media correlation learning. Thus, we propose to fully exploit multi-level cross-media alignment, which can learn more precise correlation between different media types.

\section{Our CRAN approach}

As shown in Figure \ref{fig:network}, we construct cross-media relation attention network to explore multi-level alignment, which contains three subnetworks for global, local and relation alignments respectively. 
Specifically, we utilize attention mechanism to exploit not only the local fine-grained patches, but also the relations between them. Multi-level alignment is proposed for mutually boosting, which can learn complementary hints for cross-media correlation learning. 
Then, we introduce the formal definition of cross-media dataset as $D=\left \{I,T\right \}$, where $I=\{i_p\}_{p=1}^{N}$ and text $T=\{t_q\}_{q=1}^{N}$ have totally $N$ instances in each media type. $i_p$ and $t_q$ are the $p$-th and $q$-th instance of image and text respectively. Finally, given a query of any media type, the goal of cross-media retrieval is to measure cross-media similarity $sim(i_p,t_q)$, and retrieve relevant instances of another media type.

\subsection{Visual-language relation attention model}

We extract global, local and relation representations from the proposed visual-language relation attention model, 
which can provide abundant hints for cross-media correlation learning.

For \textbf{global representation}, each input image $i_p$ is resized to $256\times 256$, and fed into a convolutional neural network to exploit high-level global semantic information. Specifically, the convolutional neural network has the same configuration with 19-layer VGGNet \cite{DBLP:journals/corr/SimonyanZ14a}, which is pre-trained on the large-scale ImageNet dataset. We extract 4,096 dimensional feature vector from fc7 layer as image global representation, denoted as $g^{i}$. 
Then, each input text $t_q$ is composed as a character sequence, where each character is represented by one-hot encoding. Following \cite{DBLP:conf/nips/ZhangZL15}, we construct a fast convolutional network for text (Char-CNN) to generate a sequence of representation from the last activation layer, and feed them into recurrent neural network. 
Specifically, we utilize long short-term memory (LSTM) network to learn global representation. The LSTM is updated recursively with the following equations.
\begin{align}
\begin{Bmatrix} i_t\\ f_t\\ o_t \end{Bmatrix}=\sigma &\left ( \begin{Bmatrix} W_i\\ W_f\\ W_o \end{Bmatrix}x_t+ \begin{Bmatrix} U_i\\ U_f\\ U_o \end{Bmatrix}h_{t-1}+\begin{Bmatrix} b_i\\ b_f\\ b_o \end{Bmatrix}\right ) \label{l1}\\
c_t=c_{t-1}\odot &f_t+\textrm{tanh}(W_ux_t+U_uh_{t-1}+b_u)\odot i_t \\
&h_t=o_t\odot \textrm{tanh}(c_t) \label{l2}
\end{align}
where the activation vectors of input, forget, memory cell and output are denoted as $i,f,c$ and $o$ respectively. $x$ is the input text sequences. Outputs from hidden units are $H_g=\{h^g_1,...,h^g_m\}$. $\odot$ denotes element-wise multiplication. $\sigma$ is sigmoid nonlinearity to activate the gate. Thus, the global representation for text can be obtained from LSTM as $g^{t}=1/m\sum_{k=1}^{m}h^g_k$.

For \textbf{local representation}, we first utilize Faster RCNN \cite{DBLP:conf/nips/RenHGS15} to generate candidate image regions, which have larger probabilities to contain visual objects, such as ``person'' or ``car''. Specifically, each image $i_p$ is fed into Faster RCNN implemented with VGG-16 network, which is pre-trained on MS-COCO detection dataset \cite{DBLP:conf/eccv/LinMBHPRDZ14}. We can obtain several bounding boxes, then we extract visual feature for each image region from the fc7 layer. They form the image local representations $\{l^{i}_1,...,l^{i}_n\}$ for $n$ different regions within one image. 
Then, for learning text local representation, we also construct Char-CNN following with LSTM, and we can obtain a sequence of outputs from the hidden units of LSTM, 
denoted as $H_l=\{h^l_1,...,h^l_m\}$ for $m$ different text fragments. 
Furthermore, we aim to make model focus on necessary fine-grained patches, so we apply attention mechanism to capture useful textual fragments.
The attention weights are calculated by a feed-forward network with softmax function as follows. 
\begin{align}
M^l=&\ \textrm{tanh}(W_a^lH_l) \label{r1}\\
a^l=&\ \textrm{softmax}(w_{la}^{\top}M^l) \label{r2}
\end{align}
where $a^l$ denotes the generated attention weights for text fragments. The fragment with larger attention weight is more likely to contain some key words, which describe the corresponding visual objects. 
Therefore, we can obtain text local representation as $l^{t}=1/m\sum_{k=1}^{m}a^l_kh^l_k$. 
It contains rich fine-grained local information, which can emphasize all key words along the text sequence. 


For the \textbf{relation representation}, 
we aim to fully model the relations between image regions as well as their language expressions. For the image relation, we utilize the image regions that are extracted from Faster RCNN as mentioned above, and we construct pairwise combinations between the regions in one image to consider their relations. Thus, the image relation representation is denoted as $r^{i}=\{l^{i}_j;l^{i}_k\}\ \ \textrm{for}\  j,k=1,...,n$, where $\{\cdot ;\cdot \}$ means the concatenation of $j$-th and $k$-th local image representations for the corresponding regions.
Then, for text relation, we apply relation attention model to allow the network focus on the relation expressions in textual description. Specifically, we also construct Char-CNN with LSTM to generate a sequence of text fragment as $H_r=\{h^r_1,...,h^r_m\}$. 
Then, the relation attention model is adopted on the top of output sequence from LSTM, which consists of a feed-forward network with softmax function with similar equations (\ref{r1}) and (\ref{r2}). We can calculate the attention weights $a^r$ for different text fragments, where the text fragments with larger relation attention weights have higher probabilities to contain the relation expressions, such as ``above'' or ``next to'', which represent the relation between key words.
Thus, we can finally generate the text relation representation as $r^{t}=\frac{1}{m}\sum_{k=1}^{m}a^r_kh^r_k$,
where the output vectors from LSTM are multiplied by the learned relation weights, aiming to enhance the textual relation information in text description.

\subsection{Cross-media Multi-level Alignment}

Since we have obtained three kinds of representations for both image and text, namely global, local and relation representations, we learn multi-level alignment to fully exploit cross-media correlation.

For \textbf{global-level alignment}, we aim to learn pairwise cross-media correlation between global original instances of different media types, as $g^{i}$ for image and $g^{t}$ for text. We design cross-media joint embedding loss for global alignment. It takes consideration that
difference between the similarity of matched image/text pair and the similarity of mismatched pair should be as large as possible. Thus, the objective function is defined as follows.
\begin{align}
\mathcal{L}_{global}=\frac{1}{N}\sum_{n=1}^{N}l_{g1}(i_n^{+},t_n^{+},t_n^{-}) + l_{g2}(t_n^{+},i_n^{+},i_n^{-})
\end{align}
The two items in this formula are defined as:
\begin{align}
l_{g1}(i_n^{+},&t_n^{+},t_n^{-})=\notag \\&\textrm{max}(0,\alpha -d(g_n^{i+},g_n^{t+})+d(g_n^{i+},g_n^{t-})) \label{li1}\\
l_{g2}(t_n^{+},&i_n^{+},i_n^{-})=\notag \\&\textrm{max}(0,\alpha -d(g_n^{i+},g_n^{t+})+d(g_n^{i-},g_n^{t+})) \label{li2}
\end{align}
where $d(.)$ denotes dot product between image/text pair. It indicates their similarity (larger is better here). $(g_n^{i+},g_n^{t+})$ denotes the matched image/text pair, 
while $(g_n^{i+},g_n^{t-})$ and $(g_n^{i-},g_n^{t+})$ are the mismatched pairs. $\alpha$ denotes the margin parameter. $N$ is the number of triplet tuples sampled from training set. Therefore, cross-media global alignment can be fully exploited from both matched and mismatched image/text pairs.

For \textbf{local-level alignment}, we aim to find the best matching between text local representation $l^{t}$ and multiple image local representations $\{l^{i}_1,...,l^{i}_n\}$ within a pair of image and text. Specifically, for each text local representation, we select $K$ nearest neighbors (KNN) from multiple image local representations, and give the following objective function.
\begin{align}
&\mathcal{L}_{local}=\notag \\&\ \textrm{max}(0,\alpha -\frac{1}{K}\sum_{k=1}^{K}d(l^{t+},l^{i+}_k)+\frac{1}{K}\sum_{k=1}^{K}d(l^{t+},l^{i-}_k))
\end{align}
where $d(.)$ is dot product indicating their similarity (also larger is better here). While we consider that average similarity of $K$ nearest local pairs in the matched image and text should be larger, compared with that in the mismatched image/text pair, which can fully exploit the cross-media local alignment.

\begin{table*}[htb]
	
	\begin{center}
		\scalebox{0.85}{
			\begin{tabular}{|c|p{1.4cm}<{\centering}|p{1.4cm}<{\centering}|p{1.4cm}<{\centering}|p{1.4cm}<{\centering}|p{1.4cm}<{\centering}|p{1.4cm}<{\centering}|} 
				\hline
				\multirow{2}{*}{Method} & \multicolumn{3}{c|}{Image annotation} & \multicolumn{3}{c|}{Image retrieval}\\
				\cline{2-7}
				& R@1 & R@5 & R@10 & R@1 & R@5 & R@10  \\
				\hline \hline
				\textbf{Our CRAN Approach} & \textbf{0.381} & \textbf{0.708} & \textbf{0.828} & \textbf{0.381} & \textbf{0.711} & \textbf{0.826}  \\
				CCL  & 0.377 & 0.694 & 0.811 &0.373 &0.684 &0.800\\
				DCCA &0.279 &0.569 &0.682 &0.268 &0.529 &0.669\\
				Corr-AE  &0.303 &0.615 &0.740 &0.238 &0.575 &0.707\\
				Multimodal DBN  &0.064 &0.194 &0.296 &0.047 &0.151 &0.232\\
				Bimodal AE  &0.127 &0.324 &0.452 &0.110 &0.328 &0.450\\
				GMM+HGLMM  &0.350 &0.620 &0.738 &0.250 &0.527 &0.660\\
				MACC  &0.139 &0.341 &0.463 &0.353 &0.660 &0.782\\
				KCCA  &0.108 &0.281 &0.399 &0.158 &0.400 &0.543\\
				CFA  &0.192 &0.449 &0.574 &0.242 &0.566 &0.683\\
				CCA  &0.076 &0.205 &0.302 &0.091 &0.268 &0.390 \\
				\hline
				
			\end{tabular} 
		}
	\end{center}
	\caption{The performance of cross-media retrieval, which shows the recall scores of two retrieval tasks on \textbf{Flickr-30K} dataset.}
	\label{table:flickr}
\end{table*}

\begin{table*}[htb]
	
	\begin{center}
		\scalebox{0.85}{
			\begin{tabular}{|c|p{1.4cm}<{\centering}|p{1.4cm}<{\centering}|p{1.4cm}<{\centering}|p{1.4cm}<{\centering}|p{1.4cm}<{\centering}|p{1.4cm}<{\centering}|} 
				\hline
				\multirow{2}{*}{Method} & \multicolumn{3}{c|}{Image annotation} & \multicolumn{3}{c|}{Image retrieval}\\
				\cline{2-7}
				& R@1 & R@5 & R@10 & R@1 & R@5 & R@10  \\
				\hline \hline
				\textbf{Our CRAN Approach} & \textbf{0.230} & \textbf{0.520} & \textbf{0.660} & \textbf{0.211} & \textbf{0.489} & \textbf{0.645}  \\
				CCL  & 0.186 & 0.474 & 0.625 &0.196 &0.469 &0.623\\
				DCCA &0.069 &0.211 &0.318 &0.066 &0.209 &0.322\\
				Corr-AE  &0.154 &0.397 &0.532 &0.138 &0.353 &0.478\\
				Multimodal DBN  &0.054 &0.194 &0.292 &0.046 &0.155 &0.240\\
				Bimodal AE  &0.063 &0.220 &0.347 &0.054 &0.178 &0.283\\
				GMM+HGLMM  &0.173 &0.390 &0.502 &0.108 &0.283 &0.401\\
				MACC  &0.056 &0.167 &0.244 &0.155 &0.370 &0.490\\
				KCCA  &0.072 &0.202 &0.305 &0.020 &0.074 &0.122\\
				CFA  &0.086 &0.258 &0.371 &0.150 &0.381 &0.514\\
				CCA  &0.041 &0.142 &0.226 &0.041 &0.155 &0.251 \\
				\hline
				
			\end{tabular} 
		}
	\end{center}
	\caption{The performance of cross-media retrieval, which shows the recall scores of two retrieval tasks on \textbf{MS-COCO} dataset.}
	\label{table:coco}
\end{table*}

For the \textbf{relation-level alignment}, we attempt to explore the alignment between text relation representation $r^{t}$ and multiple image relation representations in $r^{i}$ within a pair of image and text. Similar to local alignment, we also adopt KNN measurement to select $K$ nearest image relation representations for each text relation representation, which aims to fully model the matched relation pairs between image and text, and ignore those misalignments. Thus, the objective function is defined as follows.
\begin{align}
&\mathcal{L}_{relation}=\notag \\&\textrm{max}(0,\alpha -\frac{1}{K}\sum_{k=1}^{K}d(r^{t+},r^{i+}_k)+\frac{1}{K}\sum_{k=1}^{K}d(r^{t+},r^{i-}_k)) \label{l_re}
\end{align}
where $d(.)$ is the dot product to indicate the relation similarity between different media types, which the relation similarity is maximized between the $K$ nearest relation pairs in the matched image and text to fully explore the cross-media relation alignment.

Finally, we design the cross-media similarity between image $i_p$ and text $t_q$ that combines multi-level alignment.
\begin{align}
sim(i_p,&t_q)=\notag \\&d(g^i,g^t)+\frac{1}{K}\sum_{k=1}^{K}d(l^{i}_k,l^{t})+\frac{1}{K}\sum_{k=1}^{K}d(r^{i}_k,r^{t})
\end{align}
Thus, it can fully exploit the complementarity across multiple alignments to capture global, local and relation information, which can boost the performance of cross-media retrieval.

\subsection{Implementation Details}

Our proposed CRAN approach is implemented by Torch. For generating image regions, all candidate image regions are detected from Faster RCNN, and ordered by their scores. The top 5 patches are picked up. 
The relation representation within each image are generated by concatenating the features of every two image regions in order. As a result, each image has 20 candidate relation representations. 
For text, each sentence is treated as a character sequence, where characters are converted into one-hot vectors. 
The length of sequence is set as 201. The sentences larger than length of 201 are truncated, while those beneath the limit are padded by zeros. 
There are three convolutional layers in Char-CNN, and the parameter combinations are (384, 4), (512, 4) and (2048, 4). The first parameter means the number of kernels, and the second refers to kernel width. The outputs of Char-CNN are processed by an LSTM network. Their output dimension is 2048. 
We use fully-connected network in each subnetwork to generate global, local and relation representations, which have 1,024 dimensions. Besides, all the margins $\alpha$ in loss functions are set to 1. We set $K=3$ for local and relation alignment in cross-media similarity measurement.

\begin{table*}[htb]
	
	\begin{center}
		\scalebox{0.85}{
			\begin{tabular}{|c|c|p{1.4cm}<{\centering}|p{1.4cm}<{\centering}|p{1.4cm}<{\centering}|p{1.4cm}<{\centering}|p{1.4cm}<{\centering}|p{1.4cm}<{\centering}|} 
				\hline
				\multirow{2}{*}{Dataset} &\multirow{2}{*}{Method} & \multicolumn{3}{c|}{Image annotation} & \multicolumn{3}{c|}{Image retrieval}\\
				\cline{3-8}
				&& R@1 & R@5 & R@10 & R@1 & R@5 & R@10  \\
				\hline \hline
				\multirow{4}{*}{Flickr-30K} &\textbf{Our CRAN Approach} & \textbf{0.381} & \textbf{0.708} & \textbf{0.828} & \textbf{0.381} & \textbf{0.711} & \textbf{0.826}  \\
				&CRAN-relation  & 0.330 & 0.642 & 0.771 &0.353 &0.659 &0.788\\
				&CRAN-local &0.332 &0.645 &0.776 &0.331 &0.675 &0.786\\
				&CRAN-baseline  &0.289 &0.601 &0.722 &0.281 &0.620 &0.752\\
				\hline
				\hline
				\multirow{4}{*}{MS-COCO} &\textbf{Our CRAN Approach} & \textbf{0.230} & \textbf{0.520} & \textbf{0.660} & \textbf{0.211} & \textbf{0.489} & \textbf{0.645} \\
				&CRAN-relation  & 0.171 & 0.435 & 0.577 &0.190 &0.446 &0.600\\
				&CRAN-local &0.207 &0.494 &0.632 &0.209 &0.484 &0.633\\
				&CRAN-baseline  &0.157 &0.410 &0.548 &0.155 &0.411 &0.562\\
				\hline
				
			\end{tabular} 
		}
	\end{center}
	\caption{Baseline comparisons for each component in our proposed CRAN approach on 2 cross-media datasets.}
	\label{table:baseline}
\end{table*}

\section{Experiment}


\subsection{Datasets}
Here we briefly introduce 2 widely-used cross-media datasets adopted in the experiment as follows.
\begin{itemize}
	\item \textbf{Flickr-30K dataset} \cite{DBLP:journals/tacl/YoungLHH14} consists of 31,784 images from Flickr.com. Each image is annotated by 5 sentences. Following \cite{peng2017ccl,DBLP:conf/cvpr/TranBC16}, there are 1,000 pairs in testing set and 1,000 pairs for validation, while the rest are for training.
	
	\item \textbf{MS-COCO dataset} \cite{DBLP:conf/eccv/LinMBHPRDZ14} contains 123,287 images and each has 5 independent annotated sentences. Following \cite{peng2017ccl,DBLP:conf/cvpr/KleinLSW15}, there are both 5,000 pairs split randomly as testing set and validation set, while the rest are training set.
\end{itemize}
\subsection{Compared Methods and Evaluation Metric}

We conduct two kinds of cross-modal retrieval tasks on Flickr-30K and MS-COCO datasets, following \cite{peng2017ccl,DBLP:conf/cvpr/KleinLSW15}.
\begin{itemize}
	\item {\textbf{Image annotation}.} Retrieving groundtruth sentences given a query image (image$\to$text).
	\item {\textbf{Image retrieval}.} Retrieving groundtruth images given a query text (text$\to$image).
\end{itemize}
We report the score of Recall@K following \cite{peng2017ccl,DBLP:conf/cvpr/KleinLSW15} as evaluation metric, which includes recall rates at top 1 result (R@1), top 5 results (R@5) and top 10 results (R@10). We compare our CRAN approach with 10 state-of-the-art methods to verify its effectiveness, including 5 traditional methods 
CCA \cite{HotelingBiometrika36RelationBetweenTwoVariates}, CFA \cite{LiMM03CFA}, KCCA \cite{DBLP:journals/neco/HardoonSS04}, MACC \cite{DBLP:conf/cvpr/TranBC16}, GMM+HGLMM \cite{DBLP:conf/cvpr/KleinLSW15}, as well as 5 deep learning based methods namely Bimodal AE \cite{ngiam32011multimodal}, Multimodal DBN \cite{srivastava2012learning}, Corr-AE \cite{feng12014cross}, DCCA \cite{DBLP:conf/cvpr/YanM15} and CCL \cite{peng2017ccl}.
Note that for fair and objective comparison purpose, feature extraction of all these compared methods is exactly following \cite{peng2017ccl}, and data of different media types can be converted to common representations with the same number of dimensions in the testing stage. Thus, cross-media similarity can be directly computed between query and any data by cosine distance measurement.

\subsection{Comparisons with State-of-the-art Methods}

In this part, we compare the accuracy of cross-media retrieval to evaluate the performance of cross-media correlation learning. 
Experimental results are shown in Tables \ref{table:flickr} and \ref{table:coco}, which include recall scores of two retrieval tasks on 2 datasets. 
Obviously, our proposed CRAN approach achieves the best retrieval accuracies. 
Among all compared methods, we can draw the following observations: 
First, deep learning based methods fail to lead ahead with traditional methods, which remain large promotion space.
Second, traditional methods benefit from CNN image feature to get better performance, and even outperform some of deep learning based methods, such as CFA and GMM+HGLMM.


Then we present in-depth experimental analysis on cross-media retrieval results. 
Compared with these traditional methods, our proposed CRAN still shows clear advantage, for the fact that traditional methods mostly learn projection matrices, and they are limited to their traditional framework. They cannot fully exploit the complex cross-media correlation.
Among deep learning based methods, Multimodal DBN has the worst accuracy, because only a joint distribution is learned over the features of different media types. Bimodal AE and Corr-AE have better accuracies because they jointly consider reconstruction information. DCCA extends traditional CCA with two separate networks to maximize correlation on them.
CCL constructs hierarchical networks to model intra-media and inter-media correlation, and achieves the best accuracy because of fine-grained modeling and multi-task learning. Compared with state-of-the-art methods, our proposed CRAN approach achieves promising improvement with following 2 reasons: 
(1) Visual-language relation attention model explores both fine-grained patches and their relations of different media types, which can exploit cross-media fine-grained local information, as well as intrinsic relation information with complementary clues for correlation learning. 
(2) Cross-media multi-level alignment explores global, local and relation alignments across different media types, which can mutually boost for more precise cross-media correlation.

\subsection{Baseline Comparisons}

We conduct baseline experiment to verify the effectiveness of each component in our proposed CRAN approach. Results are shown in Table \ref{table:baseline}, where we apply only global alignment as the baseline method denoted as ``CRAN-baseline''. Besides, ``CRAN-local'' means to add local alignment over the baseline method, and ``CRAN-relation'' means to add relation alignment over the baseline. 

From the above results, we have the following observations: 
(1) Compared with ``CRAN-baseline'' that only with global alignment, both ``CRAN-local'' and ``CRAN-relation'' achieve better retrieval accuracies. It indicates that modeling the local alignment as well as the relation alignment can further boost the cross-modal learning.
(2) Our proposed CRAN approach outperforms all of them, for the fact that fusion of multi-level alignment can fully exploit their complementary information, and learn more precise cross-media correlation.

\section{Conclusion}

In this paper, we have proposed Cross-media Relation Attention Network to explore multi-level alignment between different media types. First, visual-language relation attention model is proposed to exploit both fine-grained patches and their relations, which can capture complementary hints in cross-media correlation learning. Second, cross-media multi-level alignment is proposed to model global, local and relation alignments, which can mutually boost to learn more precise cross-media correlation. We conduct experiments on cross-modal retrieval to verify the effectiveness of our approach. In future work, we will exploit unlabeled data to perform unsupervised learning for practical applications.

\section*{Acknowledgments}

This work was supported by National Natural Science Foundation of China under Grant 61771025 and Grant 61532005.

\bibliographystyle{named}
\bibliography{ijcai16}

\end{document}